\documentclass[a4paper,oneside,12pt]{elsarticle}
\usepackage{amsthm}
\usepackage{amsmath}
\usepackage{graphicx}%
\usepackage{amsfonts}%
\usepackage{amssymb}
 \usepackage{verbatim}
 \usepackage{float}
\usepackage{rotating}
\usepackage{amsmath,amsthm,amssymb}

\begin{document}
%\tableofcontents

\title{\huge{\textbf{Risk Measures in Quantitative Finance}}\\
\large{\textbf{by\\
Sovan Mitra}}} \maketitle
\section*{\large{Abstract}}
This paper was presented and written for two seminars: a national
UK University Risk Conference and a Risk Management industry
workshop. The target audience is therefore a cross section of
Academics and industry professionals.

The current ongoing global credit crunch \footnote{``Risk comes
from not knowing what you're doing", Warren Buffett.} has
highlighted the importance of risk measurement in Finance to
companies and regulators alike. Despite risk measurement's central
importance to risk management, few papers exist reviewing them or
following their evolution from its foremost beginnings up to the
present day risk measures.

This paper reviews the most important portfolio risk measures in
Financial Mathematics, from Bernoulli (1738) to Markowitz's
Portfolio Theory, to the presently preferred risk measures such as
CVaR (conditional Value at Risk). We provide a chronological
review of the risk measures and survey less commonly known risk
measures e.g. Treynor ratio.
\\\\
\textbf{Key words}: Risk measures, coherent, risk management,
portfolios, investment.

\line(1,0){400}
%\newpage
%%%%%%%%%%%%%%%%%%%%%%%%%%%%%%%%%%%%%%%%%%%%%%%%%%%%%%%%%%%%%%%%%%%%%%%%%%%%%%%%%%%%%%%%%%%%%%%%%%%%%%%%%%%%%%%%%%%%%%%%%%%%%%
\begin{comment}
\begin{center}
\begin{tabular}{c}
\\
\\
\\
\\
\\
\\
\\
\\
{\bf \large{``Risk comes from not knowing what you're doing",}}
\\
{\bf \large{Warren Buffett.}}
\\
\end{tabular}
\end{center}

\newpage
\end{comment}
%%%%%%%%%%%%%%%%%%%%%%%%%%%%%%%%%%%%%%%%%%%%%%%%%%%%%%%%%%%%%%%%%%%%%%%%%%%%%%%%%%%%%%%%%%%%%%%%%%%%%%%%%%%%%%%%%%%%%%%%%%%%%%

\section{Introduction and Outline}
Investors are constantly faced with a trade-off between adjusting
potential returns for higher risk. However the events of the
current ongoing global credit crisis and past financial crises
(see for instance \cite{LTCM} and \cite{mitraOptiHedge}) have
demonstrated the necessity for adequate risk measurement. Poor
risk measurement can result in bankruptcies and threaten collapses
of an entire finance sector \cite{LTCM2}.

Risk measurement is vital to trading in the multi-trillion dollar
derivatives industry \cite{WHY} and insufficient risk analysis can
misprice derivatives \cite{FIG}. Additionally incorrect risk
measurement can significantly underestimate particular risk types
e.g. market risk, credit risk  etc. .

This paper reviews the most significant risk measures with a
particular focus on those practised within the Financial
Mathematics/Quantitative Finance arena. The evolution of risk
measures can be categorised into four main stages:
\begin{enumerate}
\item Pre-Markowitz risk measures;
\item Markowitz Portfolio Theory based risk measures;
\item Value at Risk and related risk measures;
\item Risk Measures based on Coherent Risk Measurement Theory.
\end{enumerate}

%\section*{Paper Outline}
The outline of the paper is as follows. The paper starts with the
first risk measures proposed; unknown to many Financial
Mathematicians, these arose \textit{prior} to Markowitz's risk
measure. We then introduce Markowitz's Portfolio Theory, which
provided the first formal model of risk measurement and
diversification. After Markowitz we discuss Markowitz related risk
measures, in particular the CAPM model (Capital Asset Pricing
Model)  and other related risk measures e.g. Treynor ratio.

We then discuss the next most significant introduction to risk
measurement: Value at Risk (VaR) and explain its benefits. This is
followed by a discussion on the first axioms on risk theory -the
coherency axioms by Artzner et al. \cite{AE}. Coherent risk
measures and copulas are discussed and finally we mention the
future directions of risk measurement.

This paper was presented and written for two seminars: a national
UK University Risk Conference and a Risk Management industry
workshop. The target audience is therefore a cross section of
Academics and industry professionals.

\section{Pre-Markowitz and Markowitz Risk Measurement Era}

\subsection{Pre-Markowitz Risk Measurement}
A risk measure $\mathcal{\rho}$ is a function mapping a
distribution of losses $\mathcal{G}$ to $\mathbb{R}$, that is
\begin{eqnarray}
\mathcal{\rho}:\mathcal{G} \longrightarrow \mathbb{R}.
\end{eqnarray}
We note that some Academics distinguish between risk and
uncertainty as first defined by Knight \cite{KN}. Knight defines
risk as randomness with known probabilities (e.g. probability of
throwing a 6 on a die) whereas uncertainty is randomness with
unknown probabilities (e.g. probability of rainy weather).
However, in Financial risk literature this distinction is rarely
made.

A particularly important aspect of risk and risk measurement is
portfolio diversification. Diversification is the concept that one
can reduce total risk without sacrificing possible returns by
investing in more than one asset. This is possible because not all
risks affect all assets, for example, a new Government regulation
on phone call charges would affect the telecoms sector and perhaps
a few others deriving significant revenues or costs from phone
calls, but not every sector or company. Thus by investing in more
than 1 asset, one is less exposed to such ``asset specific" risks.
Note that there also exist risks that cannot be mitigated by
diversification, for example a rise in interest rates would affect
all businesses as they all save or spend money. We can
conceptually categorise all risks into diversifiable and
non-diversifiable risk (also known as ``systematic risk'' or
``market risk'').

Contrary to popular opinion, risk measurement and diversification
had been investigated prior to Markowitz's Portfolio Theory (MPT).
Bernoulli in 1738 \cite{bernoulli1954ent} discusses the famous St.
Petersburg paradox and that risky decisions can be assessed on the
basis of expected utility. Rubinstein \cite{RUBMPT} states
Bernoulli recognised diversification; that investing in a
portfolio of assets reduces risk without decreasing return.

Prior to Markowitz a number of Economists had used variance as a
measure of portfolio risk. For example Irving Fisher
\cite{FISHERNat} in 1906 suggested measuring Economic risk by
variance, Tobin \cite{tobin1958lpb} related investment portfolio
risks to variance of returns.

Before significant contributions were made in Financial
Mathematics to risk measurement theory, risk measurement was
primarily a securities analysis based topic. Furthermore,
securities analysis in itself was still in its infancy in the
first half of the twentieth century. Benjamin Graham, widely
considered the father of modern securities analysis, proposed the
idea of margin of safety as a measure of risk in
\cite{graham2003ii}. Graham also recommended portfolio
diversification to reduce risks. Graham's methodology of investing
(widely known as value investing) has not been pursued by the
Financial Mathematics community, partly because of its reliance of
securities rather than mathematical analysis. Exponents of the
value based investment methodology include renowned investors
Jeremy Grantham, Warren Buffett \cite{hagstrom2005wbw} and Walter
Schloss.

\subsection{Markowitz's Portfolio Theory (MPT)}
Although Financial Analysts and Economists were aware of risk,
prior to Markowitz's risk measure it was more concerned with
standard financial statement analysis, following a similar line of
enquiry to Graham \cite{graham2003ii}. However Markowitz
(\cite{markowitz1952ps}, \cite{markowitz1991pse}) was the first to
\textit{formalise} portfolio risk, diversification and asset
selection in a  mathematically consistent framework. All that was
needed were asset return means, variances and covariances. In this
respect MPT was a significant innovation in risk measurement, for
which Markowitz won the Nobel prize.

Markowitz proposed a portfolio's risk is equal to the variance of
the portfolio's returns. If we define the weighted expected return
of a portfolio $\mu_{p}$ as
\begin{eqnarray}
\mu_{p}=\displaystyle\sum_{i=1}^{N}w_{i}\mu_{i},
\end{eqnarray}
then the portfolio's variance $\sigma_{p}^{2}$  is
\begin{eqnarray}\label{MPT portrisk}
\sigma_{p}^{2}=\displaystyle\sum_{i=1}^{N}\displaystyle\sum_{j=1}^{N}\sigma
_{ij}w_{i}w_{j},
\end{eqnarray}
where
\begin{itemize}
\item N is the number of assets in a portfolio;
\item i,j are the
asset indices and $i,j \in \{1,...,N \}$ ;
\item $w_{i}$ is the asset weight, subject to the constraints:
\\\\
$0 \leq w_{i}\leq 1$,
\\\\
$\displaystyle\sum_{i=1}^{N}w_{i}=1$;
\item $\sigma _{ij}$ is the covariance of asset i with
asset j;
\item $\mu_{i}$ is the expected return for asset i.
\end{itemize}

Markowitz's portfolio theory was the first to explicitly account
for portfolio diversification as the correlation (or covariance)
between assets. From equation \ref{MPT portrisk} one observes that
$\sigma_{p}^{2}$ decreases as $\sigma _{ij}$ without necessarily
reducing $\mu_{p}$. The MPT also introduced the idea of optimising
portfolio selection by selecting assets lying on an efficient
frontier. The efficient frontier is found by minimising risk
($\sigma_{p}^{2}$) by adjusting $w_{i}$ subject to the constraint
$\mu_{p}$ is fixed; hence such portfolios provide the best
$\mu_{p}$ for minimal risk \cite{markowitz1991fpt}. Additionally,
it can be shown that the efficient frontier follows a concave
relation between $\mu_{p}$ and $\sigma_{p}^{2}$. This reflects the
idea of expected utility concavely increasing with risk. Most
portfolio managers apply a MPT framework to optimise portfolio
selection \cite{RUBMPT}.

%\subsection*{Extensions to Markowitz's Theory: Sharpe and Sortino Ratio }
Based on MPT portfolio risk measurement, Sharpe \cite{SHARPE}
invented the Sharpe Ratio $\mathcal{S}$:
\begin{eqnarray}
\mathcal{S}=\frac{\mu_{p}-R_{f}}{\sigma _{p}},
\end{eqnarray}
where $R_{f}$ is the risk-free rate of return. Sharpe's ratio can
be intepretted as the excess return above the risk free rate per
unit of risk, where risk is measured by MPT. The Sharpe ratio
provides a portfolio risk measure in terms of determining the
\textit{quality} of the portfolio's return at a given level of
risk. It is worth noting the Sharpe ratio's similarity to the
t-statistic. A discussion on the Sharpe ratio can be found at
Sharpe's website: www.stanford.edu/$\sim$wfsharpe/.

A variant on the Sharpe ratio is the Sortino Ratio
\cite{sortino1994pmd}, where we replace the denominator by the
standard deviation of the portfolio returns below $\mu_{p}$. This
ratio essentially performs the same measurement as the Sharpe
ratio but does not penalise portfolio performance for returns
above $\mu_{p}$.

It is worth mentioning that Roy \cite{ROYSF} formulated
Markowtiz's Portfolio Theory at the same time as Markowitz. As
Markowitz says \cite{RUBMPT}:\textit{``On the basis of Markowitz
(1952), I am often called the father of modern portfolio theory
(MPT), but Roy (1952) can claim an equal share of this honor."}

\subsection{Capital Asset Pricing Model (CAPM)}
The MPT in the 1960s' was computationally infeasible; it required
covariance calculations for all the assets where $N \geq 100$.
This motivated another risk measurement technique by Sharpe
\cite{SHARCapm} called CAPM, which was based on the MPT risk
model:
\begin{eqnarray}
    \mu_{i}&=& R_{f}+\beta_{i}(\mu_{m} -R_{f}),\\
    \beta_{i} &=& \dfrac{\sigma_{im}}{\sigma_{m}},
\end{eqnarray}
where
\begin{itemize}
\item $R_{f}$ is the risk-free rate of return;
\item $\mu_{m}$ is the expected market return;
\item $\beta _{i}$ is known as the beta coefficient for asset i;
\item $\sigma_{im}$ is the covariance of asset i and the market;
\item $\sigma_{m}$ is the standard deviation of the market.
\end{itemize}

The $\beta _{i}$ measures the sensitivity of the asset i's returns
to the market; a high $\beta _{i}$ implies asset i's returns
increase with the market. In the CAPM model the term $(\mu_{m}
-R_{f})$ is the market risk premium, which is the return awarded
above the risk-free rate for investing in a risky asset.

%\subsection*{CAPM and Portfolio Risk Measurement}
The CAPM theory postulates that all investors of different risk
aversion would all hold the same portfolio. This portfolio would
be a mixture of riskless and risky assets, weighted according to
the asset's market capital (number of shares outstanding $\times$
share price). Thus CAPM theory essentially suggests investors
would hold an index tracker fund and encouraged the development of
index funds. Index funds have been pioneered by  investment
managers such as John Bogle \cite{bogle1993bmf} (e.g. Vanguard 500
Index Fund).

The CAPM theory gave portfolio fund managers the first ``standard"
portfolio performance benchmark by measuring against an index's
performace. Benchmark examples are given from \cite{BENCH}:
\\\\
\normalsize{
\begin{tabular}
[c]{|l|l|}\hline \textbf{Index} & \textbf{Portfolio Benchmark}
\\\hline S \& P 500 & Large Market Capital Equity Funds\\\hline S
\& P 400 & Mid-Capital Equity Funds\\\hline Russell 2000 & Small
Capital Equity Funds\\\hline NASDAQ Composite & Technology Sector
Equity Funds\\\hline
\end{tabular}}
\\\\
\\
%\subsection*{Jensen's Portfolio Risk Measure} Based on CAPM,
Variations on the CAPM model include Jensen's risk measure. Jensen
quantifies portfolio returns above that predicted by CAPM with
$\alpha$:
\begin{eqnarray}
    \alpha=\mu_{p}-[R_{f}+\beta_{p}(\mu_{m} -R_{f})],
\end{eqnarray}
where $\beta_{p}$ is the portfolio's beta. The term $\alpha$ can
be interpretted as a measure of a portfolio manager's investment
ability or  ``beating the market".

Finally, another lesser well known portfolio ratio measure is the
Treynor ratio \cite{hubner2005gtr} $\mathcal{T}$:
\begin{eqnarray}
\mathcal{T}=\frac{\mu_{p}-R_{f}}{\beta_{p}}.
\end{eqnarray}
Similar to the Sharpe ratio, the Treynor ratio can be interpretted
as the ``quality" of portfolio return  above $R_{f}$ per unit of
risk but with risk measured on a CAPM  basis.

\section{Value at Risk (VaR)}

\subsection{VaR Risk Measure}
The next era of risk measurement after MPT can be traced to the
introduction of Value at Risk (VaR). This represented a
significant change in direction of risk measurement for the
following reasons:
\begin{itemize}
\item firstly, VaR initiated a shift in focus of using risk measures for the \textit{management} of risk in an industry context. In 1994 JP
Morgan created the VaR  risk measure, apparently to measure risk
across the whole institution under one holistic risk measure
\cite{dowd2002imr}. Previous risk measures did not focus on such a
holistic approach to risk measurement or management.

\item secondly, the Basel Committee on Banking Supervision, which standardises
international banking regulations and practises, stipulated a
market risk capital requirement based upon VaR in 1995. This
factor has subsequently fuelled interest in VaR and VaR related
measures as well as becoming a popular risk measure \cite{IMPACT}.

\item finally, previous measures focussed on explaining the return on an
asset based on some theoretical model of the risk and return
relation e.g. CAPM. VaR on the other hand shifted the focus to
measuring and quantifying the risk itself and in terms of losses
(rather than expected return).

\end{itemize}

VaR's purpose is to simply address the question ``How much can one
expect to lose, with a given cumulative probability $\zeta$, for a
given time horizon T?''. VaR is therefore defined as
\cite{szego2005mr}:
\begin{eqnarray}
F(Z(T) \leq VaR) &=& \zeta,
\end{eqnarray}
where
\begin{itemize}
 \item F(.) is the cumulative probability distribution function;
 \item Z(T) is the loss. The loss Z(t) is defined by
\begin{eqnarray}
 Z(t)=S(0)-S(t),
\end{eqnarray}
where S(t) is the stock price at time t;
 \item $\zeta$ is a cumulative probability associated with threshold value
VaR, on the loss distribution of Z(t).
\end{itemize}
To give an example of VaR, a portfolio may have a VaR of
\$10,000,000, for one day, with a cumulative probability of 90\%.
This means that the portfolio can expect a maximum loss of
\$10,000,000, over one day, with a 90\% cumulative probability. An
alternative interpretation would be that the portfolio's loss
could exceed \$10,000,000, in one day, with a cumulative
probability of 100-90=10\%. Typically $\zeta $ is chosen to be
0.90, 095 and 0.99.

\subsection{VaR Implementation}
The measurement of VaR on a portfolio presents a theoretical and
computational challenge as it is difficult to model the evolution
of a portfolio over time containing hundreds of assets
\cite{musiela2005mmf},\cite{HULLIntro}. Hence the implementation
of VaR is of particular interest to industry and academic
Researchers; the four main methods are \cite{dowd2002imr}:

\subsubsection*{VaR Historical Simulation}
Using a set of historical data we obtain an empirical probability
distribution of losses for the portfolio. One can then determine
the appropriate VaR by extracting it from the associated quantile.
Such an approach is convenient to implement and can be improved by
a range of statistical methods. For example one can apply
bootstrapping \cite{efron1979bma} when one has a small data set or
implement a weighted historical simulation approach.

\subsubsection*{VaR Parametric Approach}
The parametric approach requires an analytic solution to
determining the VaR for any cumulative probability. Unfortunately
not all distributions have solutions, however one can apply
Extreme Value Theory (Peak over Threshold and Generalised Extreme
Value distributions) to estimate VaR; the reader is referred to
\cite{cont2004fmj} for more information.

\subsubsection*{VaR Monte Carlo Simulation}
Monte Carlo simulation is a generic method of simulating some
random process (e.g. stochastic differential equation)
representing the assets or the portfolio itself. Consequently
after sufficient simulations we can obtain a loss distribution and
therefore extract VaR for different probabilities as was done for
historical simulation. One can improve Monte Carlo simulation
through various computational techniques \cite{glasserman2004mcm}
such as importance sampling, stratified sampling and antithetic
sampling.

\subsubsection*{VaR Variance-Covariance Method (also known as the Delta-Normal Method)}
Under the variance-covariance method,  we model the portfolio's
loss distribution by making two assumptions:
\begin{itemize}
\item the portfolio is linear: the change in
the portfolio's price V(t) is linearly dependent on its
constituent asset prices $S_{i}(t)$. In other words:
\begin{eqnarray}
\Delta V(t)= \displaystyle\sum_{i=1}^{{N}}\Delta S_{i}(t) .
\end{eqnarray}
A portfolio will be linear if it contains no derivatives.
Furthermore, in practise some modellers assume non-linear
portfolios are linear for analytical tractability. This assumption
is made in \cite{rockafellar2000ocv}.

\item the constituent assets have a joint
Normal return distribution, which implies the portfolio's returns
are Normally distributed. It is worth noting that the sum of
Normally distributed functions is not strictly always Normal; the
specific property of joint Normality however guarantees the
portfolio's return is Normal. Hence the linear portfolio
assumption alone cannot guarantee the portfolio's return is
Normal.

\end{itemize}
Given the  two assumptions enables us to describe the portfolio's
loss using a  Normal distribution, for which numerous analytical
equations and distribution fitting methods exist. Therefore VaR
calculation and implementation becomes significantly simpler.

\section{Coherent Risk Measures}

\subsection{Axioms of Risk Measurement}
A significant milestone in risk measurement was achieved when
Artzner et al. \cite{AE} proposed the first axioms of risk
measurement; risk measures that obeyed such axioms were called
coherent risk measures. The coherency axioms had far reaching
implications as it was no longer possible to arbitrarily assign a
function for risk measurement unless it obeyed these axioms,
consequently VaR was no longer considered an adequate risk
measure.

%\subsection{Definition of Coherent Risk Measurement and the 4 Axioms}
We now define a coherent risk measure $\rho(.)$.  Let X and Y
denote the future loss of two portfolios, then a risk measure
$\rho$ is coherent if it adheres to the four axioms:
\begin{enumerate}
\item risk is monotonic: if X $\leq $Y then $\rho $(X) $\leq \rho
$(Y);
\item risk is homogeneous: $\rho (\mathcal{\lambda} X)=\mathcal{\lambda}\rho (X)$ for $\mathcal{\lambda}>0$;

\item riskless translation invariance: $\rho(X+\chi)=\rho(X)-\chi$, where $\chi$ is a riskless bond;

\item risk is sub-additive: $\rho $(X+Y) $\leq \rho $(X)+$\rho
$(Y).
\end{enumerate}

We will now explain each axiom in turn. The monotonicity axiom
tells us that we associate higher risk with higher loss. The
homogeneity axiom ensures that we cannot increase or decrease risk
by investing differing amounts in the same stock; in other words
the risk arises from the stock itself and is not a function of the
quantity purchased\footnote{Note: this assumes we have no
liquidity risks. In reality this is not true, particularly during
the current global credit crisis.}.

The translation invariance axiom can be explained by the fact that
the investment in a  riskless bond bears no loss with probability
1. Hence we must always receive the initial amount invested. The
initial investment is subtracted because risk measures measure
loss as a positive amount, a hence gain is negative.

The subadditivity is the most important axiom because it ensures
that a coherent risk measure takes into portfolio diversification.
The axioms shows that investing in both portfolio X \textit{and} Y
results in a lower risk overall than the sum of the risks in
investing in portfolio X plus the risk in  portfolio Y separately.
VaR is not a coherent risk measure because it does not obey the
subadditivity axiom, consequently it can result in higher risk
arising from diversification.

We say a risk measure is weakly coherent if it is convex,
translationally invariant and homogeneous. It is also worth
mentioning that coherency axioms ensure the risk measure is convex
and so is amenable to computational optimisation; for more
information the reader is referred to \cite{rockafellar2000ocv}.
VaR on the other hand is non-convex and so possesses many local
minima.

\subsection{Coherent Risk Measures}
Given the introduction of coherency axioms and conclusion that VaR
was not coherent, new risk coherent measures were proposed to
capture the advantages of VaR. In particular, there was a need for
a ``holistic" risk measure and one that was simple to grasp in
that it could capture all the key risk information with three
pieces of information: probability, loss and time horizon.

In response to a coherent equivalent to VaR, a variety of VaR
related risk measures were proposed. Examples include TVaR
 (tail value at risk) \cite{artzner2003cmr}, WCE  (Worst
conditional expectation) \cite{inoue2003wce} and CVaR (conditional
value at risk).

CVaR has become a particularly popular risk measure due to its
similarity to VaR but also it assesses ``how bad things can get"
if the VaR loss is exceeded. CVaR is the expected loss given that
the VaR loss is exceeded; it is defined by:
\begin{eqnarray}
CVaR=E[Z(T)\mid Z(T)>VaR].
\end{eqnarray}
An alternative definition of CVaR is the mean of the tail
distribution of the VaR losses. An additional advantage of CVaR is
that the portfolio weights can be easily optimised by linear
programming \cite{rockafellar2000ocv} to minimise CVaR.

Spectral risk measures are a group of coherent risk measures,
whereby the risk is given as the sum of a weighted average of
outcomes. The weights can be chosen to reflect risk preferences
towards particular outcomes. For more information the reader is
referred to \cite{acerbi2002smr}.

\subsection{Copulas}
The subaddivity axiom demonstrates the importance to capture
dependencies between stocks when measuring the risk of a
portfolio. Consequently this gave rise to the interest in copulas
\cite{nelsen2006ic}; these are functions mapping a set of marginal
distributions into a multivariate distribution and vice versa.
Sklar's Theorem underpins copula theory, which states that for a
given multivariate distribution there exists a copula that can
combine all the marginal distributions to give the joint
distribution. For example, in the bivariate case if we have two
marginal distributions F(x) and G(y) then there exists a copula
function $\mathcal{C}$ to give the multivariate distribution
H(x,y):
\begin{eqnarray}
H(x,y)=\mathcal{C}(F(x),G(y)).
\end{eqnarray}

There exist a variety of copulas and examples include the Gaussian
copula \cite{frey2001cac} and Clayton copula
\cite{cuvelier2005cca}. Prior to their application in Financial
Mathematics copulas have been used for many years in Actuarial
Sciences, Reliability Engineering, Civil and Mechanical
Engineering.

In Extreme Value Theory copulas become extremely important because
it is not possible to capture dependencies between random
variables using standard correlation. To use multivariate Extreme
Value Theory, instead we must apply copulas to capture
dependencies \cite{dowd2002imr}.

Despite the number of copulas in existence this continues to be an
area of active research as it is important to have copulas that
capture the correct type of dependencies between stocks. For
instance copulas are used in the pricing of collateralized debt
obligations (CDOs) \cite{meneguzzo2004csc}. However the usage of
credit derivatives have been widely cited as an important cause of
the current global credit crisis.

\section{Future Directions of Risk Measurement}
Risk measurement is a thriving area of research. A current area of
interest is to find satisfactory methods of modelling dependencies
between stocks other than by copulas and correlations.
Alternatively, there is much interest in finding copulas that can
meaningfully capture dependency behaviours.

Another area of risk measurement research is dynamic risk
measurement. This involves measuring risk in continuous time,
rather than applied to a static distribution. Examples of dynamic
risk measurement include \cite{riedel2004dcr},\cite{geman2008tcm},
\cite{detlefsen2005cad}.

Finally, another  future direction of risk measurement is devising
risk specific risk measures, such credit risk measures, liquidity
risk measure etc. . However it should be noted that such risk
measures already exist in these areas (for example Merton's
structural credit default model \cite{merton1974pcd} and the KMV
model \cite{kealhofer2003qcr}).

\section{Conclusions}
This paper has surveyed the key risk measures in Financial
Mathematics as well as its progressive development since the
beginning. We have also mentioned, contrary to popular knowledge,
that risk measures existed prior to Markowitz.

We have examined the key contributions of the major risk measures,
such as Markowitz's Portfolio Theory, whilst also highlighting
their influence within the financial industry. We have also
discussed newer risk measures such as  spectral risk measures, VaR
and its variants (e.g. CVaR) and mentioned future areas of
research.

\newpage
\bibliographystyle{plain}
\addcontentsline{toc}{section}{References}
\bibliography{Ref}

\end{document}